\documentclass[11pt,a4paper]{article}

\usepackage[utf8]{inputenc}
\usepackage[T1]{fontenc}
\usepackage{lmodern}
\usepackage[english]{babel}
\usepackage{geometry}
\usepackage{setspace}
\usepackage{microtype}
\usepackage{amsmath,amssymb}
\usepackage{siunitx}
\usepackage{booktabs}
\usepackage{caption}
\usepackage{textcomp}
\usepackage{hyperref}
\hypersetup{colorlinks=true,linkcolor=black,citecolor=black,urlcolor=blue}
\usepackage{indentfirst}
\usepackage{enumitem}
\setlist[itemize]{leftmargin=*,itemsep=2pt,parsep=0pt,topsep=4pt}
\usepackage{titlesec}
\titleformat{\section}
  {\normalfont\Large\bfseries}
  {\thesection}
  {1em}
  {}
\titlespacing*{\section}
  {0pt}{3.0ex plus 1ex minus .2ex}{1.5ex plus .2ex}
\titleformat{\subsection}
  {\normalfont\large\bfseries}
  {\thesubsection}
  {1em}
  {}
\titlespacing*{\subsection}
  {0pt}{2.0ex plus .5ex minus .2ex}{1ex plus .2ex}
\usepackage{graphicx}
\usepackage{float}

\begin{document}

\title{\textbf{Radiation Effects on Breakdown Voltage and Quenching Resistance in SiPMs}}
\author{Matteo Veratelli \and Wander Baldini}
\date{}
\maketitle

\noindent\textbf{Abstract.}\enspace This study investigates the suitability of the Hamamatsu S13360 series as replacement photodetectors for the RICH detector in LHCb Upgrade II. I--V curves enabled a statistical analysis of breakdown voltage and quenching resistance before and after irradiation, quantifying the dependence of damage on neutron fluence and the mitigating effect of cooling. The results show that breakdown voltage degradation increases with neutron fluence and becomes statistically significant only at high irradiation levels. Quenching resistance, by contrast, shows only a limited increase except at $6 \times 10^{13}$\,neq/cm$^2$, where substantial degradation appears. Both parameters exhibit a clear reduction in radiation-damage impact when operated at cryogenic temperatures.

%\tableofcontents

\section{Introduction}

\subsection{LHCb and RICH detector}

LHCb (Large Hadron Collider beauty) is one of the four major experiments operating at CERN's Large Hadron Collider (LHC). Its primary objective is to improve our understanding of CP symmetry violation and to investigate the origin of the matter--antimatter asymmetry in the Universe. By studying the decay of particles containing bottom (b) quarks, LHCb aims to explain why matter was not completely annihilated after the Big Bang, thereby allowing the Universe to evolve into its present form.~\cite{ref1}

To identify the particles produced in the collisions, LHCb employs several specialized sub-detectors. Among these, two Ring Imaging Cherenkov (RICH) detectors distinguish between pions, kaons, and protons. Travelling charged particles emit cones of Cherenkov light, whose opening angle depends on the particle velocity. The RICH detectors then measure this angle and by knowing their momentum the particle mass can be determined, enabling reliable particle identification.

To study the light cones the Cherenkov photons are detected by photodetectors, which convert them into electrical signals and measure their position.

\subsection{SiPM}

The importance of high-performance photodetectors is therefore evident. To meet such new demands, LHCb is planning a major detector upgrade, known as Upgrade II, which will allow operation at substantially higher instantaneous luminosities (about five times greater) starting in 2035. The upgraded RICH detector must provide performance comparable to --- or better than --- the current system while operating under significantly more demanding radiation conditions. Achieving these goals requires replacing the existing photodetectors, Multi-Anode Photo-Multipliers (MAPMTs), with Silicon Photomultipliers (SiPMs) emerging as the most promising candidate technology.~\cite{ref2}

\begin{figure}[t]
  \centering
  \includegraphics[width=0.85\linewidth]{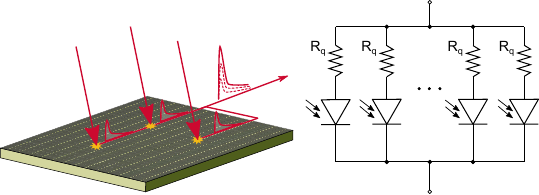}
  \caption{Left: schematic of a SiPM showing multiple microcells triggered by incident photons, whose signals sum at the output. Right: equivalent circuit of the SiPM array with all microcells connected in parallel, enabling photon counting. Adapted from~\cite{ref4}.}
  \label{fig:sipm_schematic}
\end{figure}

SiPMs consist of arrays of microcells, each composed of a Single Photon Avalanche Diode (SPAD), connected in series with a quenching resistor and fabricated on a silicon substrate as shown in Figure~\ref{fig:sipm_schematic}. The silicon is appropriately doped to create p--n junctions that are then reverse-biased. Under normal operating conditions, only a very small leakage current flows (in the order of nA), but when the applied voltage exceeds the breakdown voltage, an avalanche multiplication process is triggered, producing a measurable electrical signal. When this happens, the SPAD is said to operate in Geiger mode. The avalanche process is terminated when the voltage drop in the quenching resistor reduces the bias voltage below the breakdown threshold, thereby restoring the microcell to its initial state.~\cite{ref3}

Compared with conventional vacuum photomultiplier tubes (PMTs), SiPMs offer several important advantages. They are significantly more compact, benefit from the mature and cost-effective silicon semiconductor manufacturing industry, require much lower operating voltages, and can operate inside magnetic fields because charge carriers move through solid silicon rather than vacuum. Furthermore, SiPMs exhibit higher photon detection efficiency (PDE), and their SPAD parallel array structure allows the output signal to be proportional to the number of photons, since sufficiently energetic photons can simultaneously trigger multiple microcells.~\cite{ref2}

The principal limitation of SiPM technology is its relatively high dark count rate (DCR), of the order of $\sim 10$--$100\,\mathrm{kHz/mm^2}$ at room temperature, which is considerably larger than that of conventional PMTs at the same temperature and is approximately proportional to the active area of the device. Consequently, large-area SiPMs are more challenging to design. Operating at lower temperatures provides an effective means of reducing the DCR, although this solution introduces additional complexity and cost. Nevertheless, the robustness and flexibility of SiPM technology make it an attractive candidate for future high-energy physics detectors.~\cite{ref2}

\section{Materials and Methods}

\subsection{Hamamatsu S13360 SiPMs}

The Hamamatsu Photonics S13360 series Silicon Photomultipliers (SiPMs), marketed as Multi-Pixel Photon Counters (MPPCs), are optoelectronic semiconductor devices characterized by excellent photon-counting capability and low operating voltage. Compared with earlier generations, the S13360 series exhibits significantly reduced optical crosstalk, dark count rate (DCR), and afterpulsing.~\cite{ref5}

For the requirements of the LHCb Upgrade II, the selected SiPMs must achieve a time resolution better than 100\,ps RMS and a dark count rate below 100\,kHz/mm$^2$ for devices with active areas between $1.5\times1.5\,\mathrm{mm^2}$ and $2\times2\,\mathrm{mm^2}$. In addition, the sensor array must provide a high fill factor, exceeding 70--80\%.~\cite{ref6}

The radiation environment expected during Upgrade II corresponds to 1\,MeV-neutron-equivalent fluences (from now on referred to as neq/cm$^2$) between $3 \times 10^{13}$ and $6 \times 10^{13}$\,neq/cm$^2$~\cite{ref7}. To mitigate radiation-induced degradation under these conditions, the SiPMs will be cooled down through liquid nitrogen. Consequently, candidate devices must exhibit stable performance at cryogenic temperatures and reduction in radiation damage of parameters such as the breakdown voltage ($V_{\mathrm{bd}}$) and the quenching resistance ($R_{\mathrm{q}}$), which constitute the focus of the present study.

For this investigation, the Hamamatsu S13360-1350PE model was selected. This device features 50\,$\mu$m microcells, an active area of $1.3\times1.3\,\mathrm{mm^2}$, and a 74\% fill factor (see Table~\ref{tab:specs}), closely matching the performance requirements of the LHCb Upgrade II.

\begin{table}[htbp]
\centering
\caption{Main specifications of the Hamamatsu S13360-1350PE SiPM.~\cite{ref5}}
\label{tab:specs}
\begin{tabular}{@{}ll@{}}
\toprule
Parameter & Value \\
\midrule
Pixel size                    & 50\,$\mu$m \\
Active area                   & $1.3 \times 1.3$\,mm$^2$ \\
Package                       & glass epoxy \\
Number of pixels ($N_{\mathrm{cell}}$) & 667 \\
$V_{\mathrm{bd}}$ at $25^\circ$C & $53 \pm 5$\,V \\
Fill factor                   & 74\% \\
\bottomrule
\end{tabular}
\end{table}

\subsection{Ferrara experimental setup}

The INFN Ferrara laboratories are equipped with a multifunctional setup for SiPM characterization over a wide temperature range (78\,K -- room temperature). The system includes:
\begin{itemize}
\item a 14\,l liquid-nitrogen cryostat;
\item a Keithley SM2450 SourceMeter Unit with current resolution $< 1$\,pA, connected through triaxial cables;
\item a Keysight E3630 stabilized power supply;
\item a Tektronix MSO64 oscilloscope with 12-bit resolution, 2.5\,GHz bandwidth, and 25\,GS/s sampling rate;
\item a software-controlled mechanical linear stage;
\item a 470\,nm LED illumination system;
\item a Hamamatsu PLP10 pulsed UV laser operating at 405\,nm with a pulse width of 50\,ps;
\item a fast amplifier developed by INFN Milano-Bicocca, designed to work at cryogenic temperatures.
\end{itemize}

This experimental setup enables the characterization of several SiPM parameters, including quenching resistance and breakdown voltage through I--V curves, dark count rate (DCR), afterpulsing, optical crosstalk, and timing resolution.

\subsection{I-V curves measurement procedure and irradiation}

The quenching resistance ($R_{\mathrm{q}}$) and breakdown voltage ($V_{\mathrm{bd}}$) values were determined by measuring the current--voltage (I--V) characteristics of 15 Hamamatsu S13360-1350PE SiPMs.

Measurements were performed at five different temperatures: 297\,K (room temperature), 223\,K, 173\,K, 123\,K, and 78\,K, with an estimated relative uncertainty of about 1\%. Adjusting the distance between the SiPM and the liquid-nitrogen level set the desired temperature. Before each measurement, the detector was completely shielded from ambient light, and sufficient time was allowed for thermal equilibrium to be reached.

\begin{figure}[t]
  \centering
  \includegraphics[width=\linewidth]{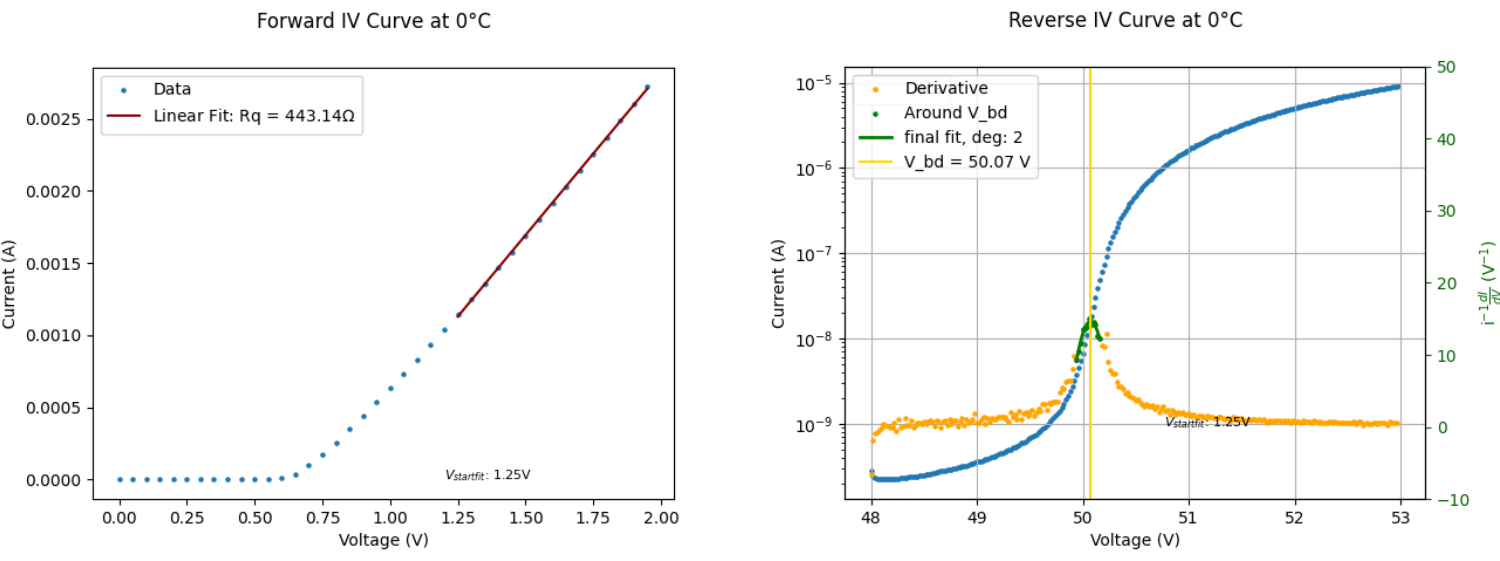}
  \caption{Representative I--V characteristics of a Hamamatsu S13360-1350PE at $0^\circ\mathrm{C}$ ($273.15\,\mathrm{K}$). Left: forward-bias curve with linear fit used to extract $R_{\mathrm{q}}$. Right: reverse-bias curve (blue) and its derivative $\mathrm{d}I/\mathrm{d}V$ (orange); the peak of the derivative identifies $V_{\mathrm{bd}}$.}
  \label{fig:iv_curves}
\end{figure}

Forward and reverse-bias I--V curves were acquired by sweeping the bias voltage ($V_{\mathrm{bias}}$) using voltage steps of 20--50\,mV (Figure~\ref{fig:iv_curves}). The applied voltages at each operating temperature are summarized in Table~\ref{tab:voltages}.

\begin{table}[htbp]
\centering
\caption{Bias voltages used for I--V measurements.}
\label{tab:voltages}
\begin{tabular}{@{}ccccc@{}}
\toprule
$T$ ($^\circ$C) & $T$ (K) & OV = 2 & OV = 3 & OV = 4 \\
\midrule
24   & 297 & 54.7 & 55.7 & 56.7 \\
$-50$  & 223 & 50.6 & 51.6 & 52.6 \\
$-100$ & 173 & 48.1 & 49.1 & 50.1 \\
$-150$ & 123 & 46.2 & 47.2 & 48.2 \\
$-195$ & 78  & 45.1 & 46.1 & 47.1 \\
\bottomrule
\end{tabular}
\end{table}

\textbf{Breakdown Voltage Extraction:} the maximum of $\mathrm{d}I/\mathrm{d}V$ on the reverse-bias curve marks the breakdown voltage, which corresponds to the inflection point where avalanche multiplication begins and causes a rapid rise in current.~\cite{ref8}

\textbf{Quenching Resistance Extraction:} At sufficiently high forward currents (in the milliampere range), the SiPM behaves as an ohmic device, resulting in a linear current--voltage relationship. A linear fit on this region provided the total device resistance; multiplying it by $N_{\mathrm{cell}} = 667$ then yielded the individual microcell $R_{\mathrm{q}}$.~\cite{ref9}

Following the initial electrical characterization, all 15 SiPMs were irradiated at the TRIGA Mark II research reactor of the Jožef Stefan Institute in Ljubljana, operating at 250\,kW. The devices were divided into five groups of three samples according to the target neutron-equivalent fluence:
\begin{itemize}
\item $3 \times 10^{11}$\,neq/cm$^2$ (devices \#101--103),
\item $3 \times 10^{12}$\,neq/cm$^2$ (devices \#104--106),
\item $6 \times 10^{12}$\,neq/cm$^2$ (devices \#107--109),
\item $3 \times 10^{13}$\,neq/cm$^2$ (devices \#110--112),
\item $6 \times 10^{13}$\,neq/cm$^2$ (devices \#113--115).
\end{itemize}

After irradiation, the devices travelled under standard shipping conditions at room temperature for approximately two days. Upon arrival at the Ferrara laboratory, they were stored at $-20\,^\circ$C until characterization. The same I--V procedure produced the post-irradiation values, which were then compared with the pre-irradiation data to quantify the radiation damage.

\section{Results}

\subsection{Pre-irradiation values}

Before irradiation, $V_{\mathrm{bd}}$ exhibited high uniformity across all devices, with a coefficient of variation below 0.4\% for every irradiation group and operating temperature. In all cases, lowering the temperature reduced $V_{\mathrm{bd}}$~\cite{ref2,ref10,ref12}.

In contrast, $R_{\mathrm{q}}$ displayed greater device-to-device and group-to-group variability, with the spread growing further at lower temperatures. In particular, devices \#110--115 exhibited the largest dispersion, with coefficients of variation ranging between 3\% and 4\% at all temperatures, whereas the devices assigned to the lowest irradiation fluences (\#101--106) showed the smallest relative variability (CV $< 2$\%).

At this stage of the study, however, none of the groups had yet undergone different treatments; therefore, the apparent association between higher initial dispersion and subsequent exposure to high irradiation levels is purely coincidental. Despite the intrinsic variability, all fifteen SiPMs exhibited the same overall trend of $R_{\mathrm{q}}$ increasing at lower temperatures~\cite{ref10,ref12,ref13}.

The average pre-irradiation values of \(V_{\mathrm{bd}}\) and \(R_{\mathrm{q}}\) are summarised in Tables~\ref{tab:vbd-pre} and~\ref{tab:rq-pre}, respectively.

\begin{table}[htbp]
\centering
\caption{Average pre-irradiation breakdown voltage.}
\label{tab:vbd-pre}
\begin{tabular}{@{}lccccc@{}}
\toprule
Group & $V_{\mathrm{bd}}$ 297\,K (V) & $V_{\mathrm{bd}}$ 223\,K (V) & $V_{\mathrm{bd}}$ 173\,K (V) & $V_{\mathrm{bd}}$ 123\,K (V) & $V_{\mathrm{bd}}$ 78\,K (V) \\
\midrule
\#101--103 & $51.83 \pm 0.08$ & $47.67 \pm 0.16$ & $45.17 \pm 0.08$ & $43.21 \pm 0.08$ & $42.08 \pm 0.06$ \\
\#104--106 & $51.66 \pm 0.13$ & $47.41 \pm 0.17$ & $44.90 \pm 0.06$ & $43.00 \pm 0.13$ & $41.96 \pm 0.11$ \\
\#107--109 & $51.69 \pm 0.08$ & $47.42 \pm 0.09$ & $44.92 \pm 0.05$ & $42.97 \pm 0.05$ & $41.91 \pm 0.06$ \\
\#110--112 & $51.73 \pm 0.04$ & $47.38 \pm 0.12$ & $44.90 \pm 0.10$ & $43.01 \pm 0.06$ & $41.94 \pm 0.06$ \\
\#113--115 & $51.79 \pm 0.07$ & $47.50 \pm 0.15$ & $45.01 \pm 0.12$ & $43.05 \pm 0.02$ & $41.98 \pm 0.06$ \\
\bottomrule
\end{tabular}
\end{table}

\begin{table}[htbp]
\centering
\caption{Average pre-irradiation quenching resistance.}
\label{tab:rq-pre}
\begin{tabular}{@{}lccccc@{}}
\toprule
Group & $R_{\mathrm{q}}$ 297\,K ($\Omega$) & $R_{\mathrm{q}}$ 223\,K ($\Omega$) & $R_{\mathrm{q}}$ 173\,K ($\Omega$) & $R_{\mathrm{q}}$ 123\,K ($\Omega$) & $R_{\mathrm{q}}$ 78\,K ($\Omega$) \\
\midrule
\#101--103 & $425 \pm 3$   & $500 \pm 6$   & $572 \pm 7$   & $677 \pm 13$  & $823 \pm 11$ \\
\#104--106 & $418.6 \pm 0.5$ & $500 \pm 5$   & $572 \pm 7$   & $678 \pm 12$  & $818 \pm 13$ \\
\#107--109 & $422 \pm 7$   & $499 \pm 8$   & $573 \pm 12$  & $684 \pm 20$  & $824 \pm 27$ \\
\#110--112 & $443 \pm 17$  & $526 \pm 21$  & $604 \pm 25$  & $718 \pm 29$  & $869 \pm 37$ \\
\#113--115 & $448 \pm 13$  & $531 \pm 19$  & $610 \pm 22$  & $727 \pm 19$  & $880 \pm 30$ \\
\bottomrule
\end{tabular}
\end{table}

\subsection{Breakdown voltage variation}

\begin{figure}[p]
  \centering
  \includegraphics[width=0.96\linewidth]{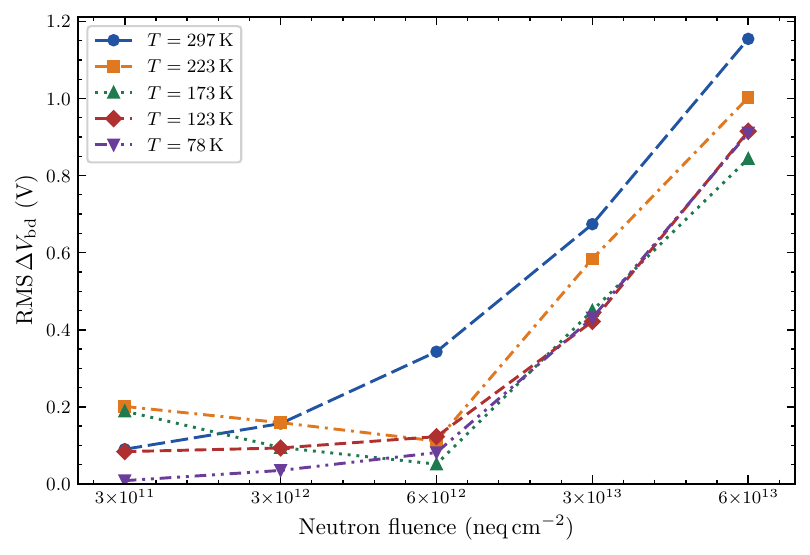}
  \caption{RMS of $\Delta V_{\mathrm{bd}}$ as a function of neutron
           fluence for all five operating temperatures. Each point
           represents the RMS over the three devices in the same
           irradiation group.}
  \label{fig:rms_dvbd_flu}

  \vspace{1.5em}

  \includegraphics[width=0.96\linewidth]{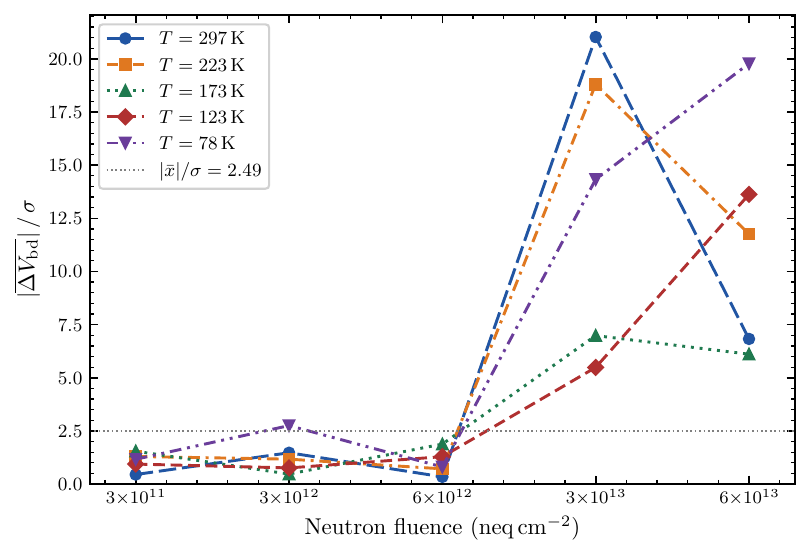}
  \caption{Ratio $|\overline{\Delta V_{\mathrm{bd}}}|/\sigma$ as a
           function of neutron fluence for all five temperatures.
           The dashed line at $|\bar{x}|/\sigma = 2.49$ marks the
           $p < 0.05$ significance threshold for $n = 3$ (df\,$ = 2$).
           Values above this line indicate that the radiation-induced
           shift exceeds the intrinsic device spread.}
  \label{fig:snr_dvbd}
\end{figure}

\textbf{Dependence on Fluence at Constant Temperature:} Following irradiation, the breakdown voltage generally increased. The variation in breakdown voltage ($\Delta V_{\mathrm{bd}}$) exhibited, in Figure~\ref{fig:rms_dvbd_flu}, an approximately exponential dependence on neutron fluence, with the most pronounced increase occurring for fluences above $6 \times 10^{12}$\,neq/cm$^2$.

However, comparison between the mean variation and the corresponding standard deviation of each irradiation group in Figure~\ref{fig:snr_dvbd} shows that, at relatively low fluences ($< 6 \times 10^{12}$\,neq/cm$^2$), the observed shifts in $V_{\mathrm{bd}}$ are negligible and comparable to the intrinsic device-to-device variability~\cite{ref13,ref14}. By contrast, devices irradiated to $3 \times 10^{13}$ and $6 \times 10^{13}$\,neq/cm$^2$ (\#110--115) exhibit $\Delta V_{\mathrm{bd}}$ values substantially exceeding the intrinsic spread, with signal-to-noise ratios ranging from approximately 6 to 21. At these higher fluences, radiation damage therefore becomes statistically significant~\cite{ref13,ref15,ref16} and largely independent of individual device characteristics, reaching increases of 0.91--1.15\,V at $6 \times 10^{13}$\,neq/cm$^2$.

\textbf{Dependence on Temperature at Constant Fluence:} As the operating temperature decreases, $\Delta V_{\mathrm{bd}}$ remains positive but gradually decreases, suggesting that cryogenic operation mitigates the impact of radiation-induced degradation. The relatively scattered values observed at room temperature are likely attributable to small variations in ambient laboratory temperature.

\begin{figure}[t]
  \centering
  \includegraphics[width=\linewidth]{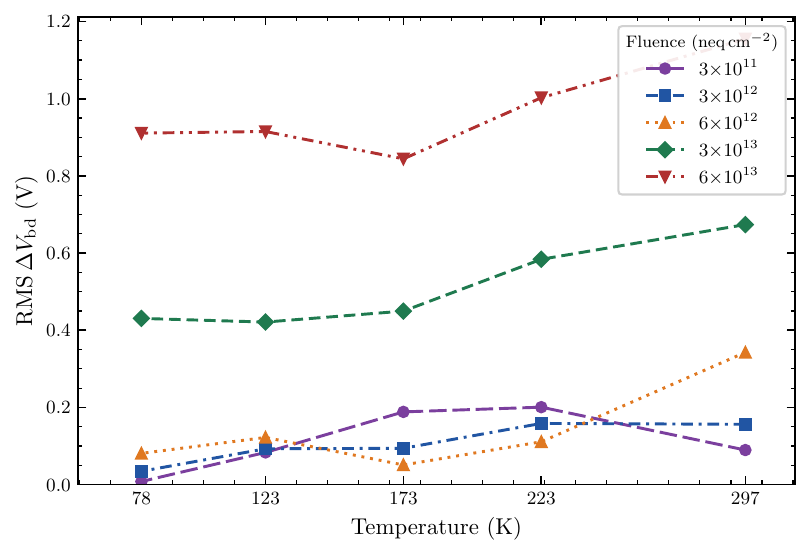}
  \caption{RMS of $\Delta V_{\mathrm{bd}}$ as a function of operating temperature for each irradiation group. The systematic decrease at lower temperatures is consistent across all fluence levels.}
  \label{fig:rms_dvbd_temp}
\end{figure}

This behavior is observed for all irradiation levels (Figure~\ref{fig:rms_dvbd_temp}). To evaluate its statistical significance, $t$-tests were performed on the $\Delta V_{\mathrm{bd}}$ values of the three devices within each irradiation group at every temperature. For the lower irradiation levels ($3 \times 10^{11}$, $3 \times 10^{12}$, and $6 \times 10^{12}$\,neq/cm$^2$), the shifts in breakdown voltage remain comparable to the intrinsic variability of the devices, resulting in relatively low statistical significance ($|t| < 4.3$, df $= 2$); although this limited significance could be a consequence of the smaller damage at these fluences.

\begin{figure}[t]
  \centering
  \includegraphics[width=\linewidth]{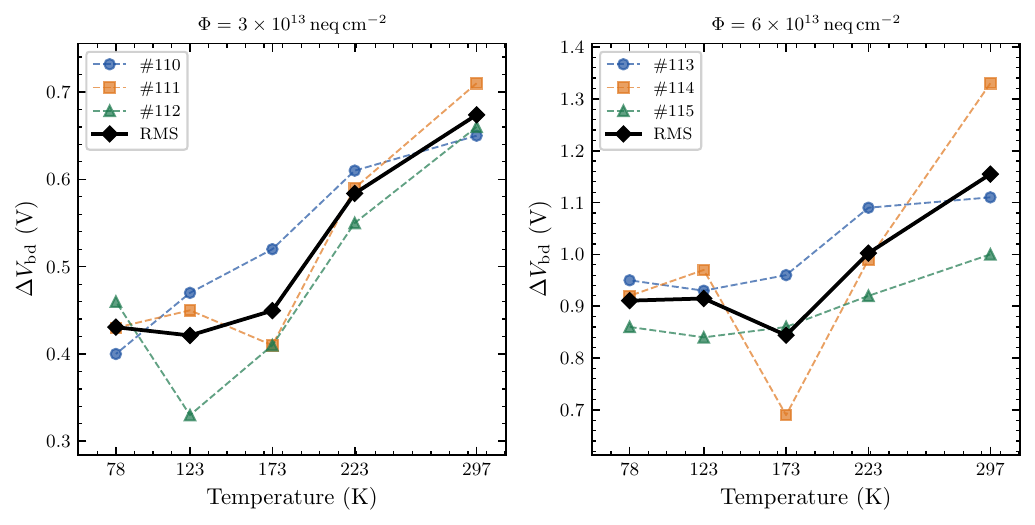}
  \caption{$\Delta V_{\mathrm{bd}}$ as a function of temperature for individual devices and group RMS at $3\times10^{13}$\,neq/cm$^2$ (left) and $6\times10^{13}$\,neq/cm$^2$ (right). Dashed lines show individual devices; the solid line shows the group RMS.}
  \label{fig:dvbd_ind}
\end{figure}

At higher irradiation levels, instead, shown in Figure~\ref{fig:dvbd_ind}, $\Delta V_{\mathrm{bd}}$ becomes substantially larger and significant ($|t| > 9.9$, with average values around 25). Under these conditions, cooling the irradiated devices from room temperature to cryogenic temperatures reduces the observed damage by approximately 20--40\%, after which $\Delta V_{\mathrm{bd}}$ stabilizes at a non-zero value.

Further evidence supporting the reduced impact of irradiation at low temperatures is provided by the thermal coefficient $\mathrm{d}V_{\mathrm{bd}}/\mathrm{d}T$, obtained by linear regression. Across all irradiation levels, this coefficient remains remarkably constant at $45.0\pm0.3\,\mathrm{mV/K}$. Thus, radiation modifies the absolute value of the breakdown voltage but does not significantly affect its temperature dependence.

\begin{table}[htbp]
\centering
\caption{Mean variation $\Delta V_{\mathrm{bd}}$ after irradiation, and thermal coefficient $\mathrm{d}V_{\mathrm{bd}}/\mathrm{d}T$.}
\label{tab:dvbd}
\begin{tabular}{@{}lcccccc@{}}
\toprule
Fluence (neq/cm$^2$) & 297\,K & 223\,K & 173\,K & 123\,K & 78\,K & $\mathrm{d}V_{\mathrm{bd}}/\mathrm{d}T$ (mV/K) \\
\midrule
$3 \times 10^{11}$ & 0.090 & 0.201 & 0.189 & 0.083 & 0.008 & 45.1 \\
$3 \times 10^{12}$ & 0.156 & 0.159 & 0.094 & 0.093 & 0.035 & 44.9 \\
$6 \times 10^{12}$ & 0.343 & 0.111 & 0.051 & 0.122 & 0.081 & 45.3 \\
$3 \times 10^{13}$ & 0.674 & 0.584 & 0.450 & 0.421 & 0.431 & 45.2 \\
$6 \times 10^{13}$ & 1.155 & 1.002 & 0.844 & 0.915 & 0.911 & 45.4 \\
\bottomrule
\end{tabular}
\end{table}

Notably, at a fluence of $3 \times 10^{11}$\,neq/cm$^2$ and 78\,K, the average $\Delta V_{\mathrm{bd}}$ is only $0.008 \pm 0.006$\,V (Table~\ref{tab:dvbd}), with individual values of $\{0.00, 0.01, 0.01$\,V$\}$, suggesting that, at sufficiently low irradiation levels, operation near liquid-nitrogen temperature may reduce radiation shifts to values below the experimental resolution of the measurement setup.

\subsection{Quenching resistance variation}

Before discussing the post-irradiation results, it is important to note that even before irradiation $R_{\mathrm{q}}$ already showed larger dispersion than $V_{\mathrm{bd}}$ (coefficients of variation up to 4\,\%). Irradiation amplified this variability further.~\cite{ref13}

To quantify this dispersion, the ratio between the data spread and the median value was examined. Except for group \#110--112, which exhibited a ratio close to 1, all other groups displayed spreads between 4 and 27 times larger than their median values, with an average factor of approximately 8. Consequently, the median, together with the RMS and arithmetic mean, was considered, as it is less sensitive to the large dispersion and the presence of outliers.

\begin{figure}[p]
  \centering
  \includegraphics[width=0.96\linewidth]{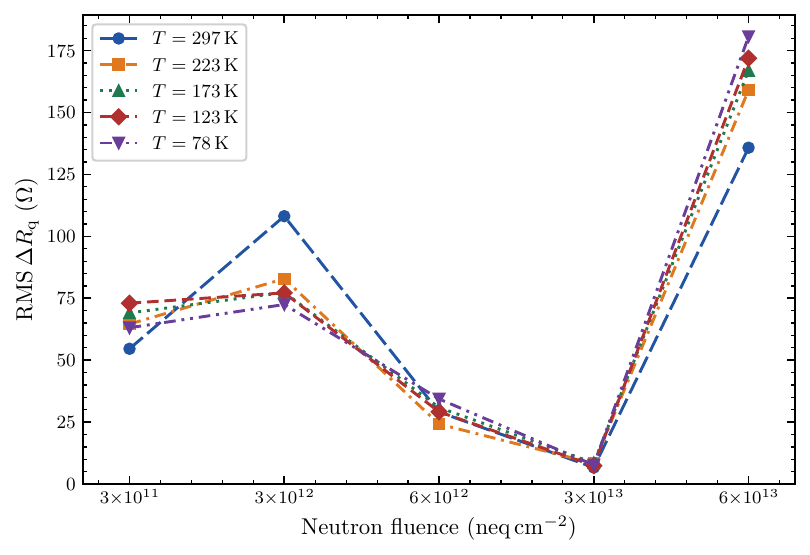}
  \caption{RMS of $\Delta R_{\mathrm{q}}$ as a function of neutron fluence at five operating temperatures. The large spread at $3\times10^{12}$ and $6\times10^{13}$\,neq/cm$^2$ reflects the dominance of high-response individual devices.}
  \label{fig:rms_drq_flu}

  \vspace{1.5em}

  \includegraphics[width=0.96\linewidth]{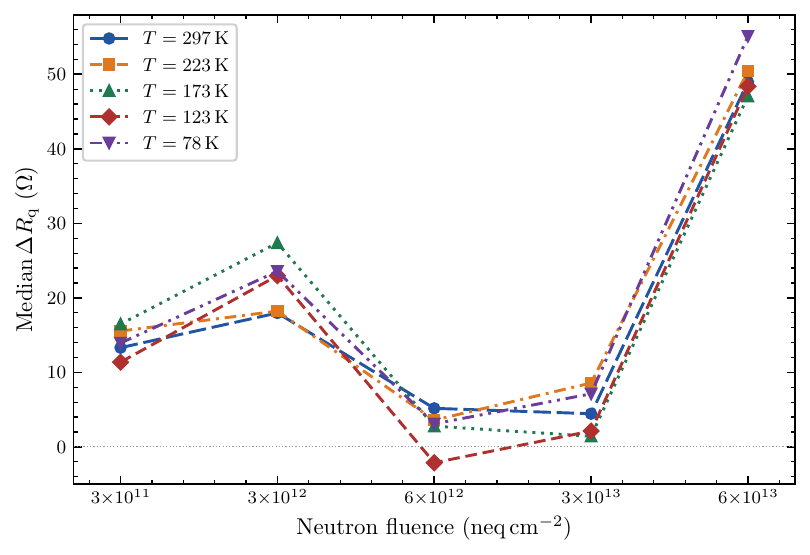}
  \caption{Median $\Delta R_{\mathrm{q}}$ as a function of neutron fluence at five operating temperatures. With $n=3$ per group, the median equals the middle-ranked device value and is more robust to the influence of extreme outliers.}
  \label{fig:med_drq_flu}
\end{figure}

\textbf{Dependence on Fluence at Constant Temperature:} Following irradiation, $R_{\mathrm{q}}$ generally increases, with only a few isolated exceptions, most notably device \#109, discussed later. Relative to $V_{\mathrm{bd}}$, the $R_{\mathrm{q}}$ response is more complex and displays far larger device-to-device scatter.

This large internal spread makes it difficult to identify a unique dependence on neutron fluence. Nevertheless, median relative values suggest limited degradation below $6 \times 10^{13}$\,neq/cm$^2$ and a clear rise at that highest fluence (see Table~\ref{tab:drq-rel}).

$\Delta R_{\mathrm{q}}$ plots, Figures~\ref{fig:rms_drq_flu} and~\ref{fig:med_drq_flu}, display an oscillatory trend. To determine whether this behavior is physical or merely statistical, a combinatorial analysis was performed for all 243 possible sequences obtained by selecting one device from each irradiation group at constant temperature. The results are reported in Table~\ref{tab:up-percent}.

The analysis showed that the UDDU (up--down--down--up) pattern visible in the plotted medians occurs in only 17--28\% of all possible sequences. Instead, the most common pattern is UDUU (up--down--up--up), more common at all temperatures and representing approximately one-third of all combinations.

\begin{table}[htbp]
\centering
\caption{Median relative variation of the quenching resistance after irradiation.}
\label{tab:drq-rel}
\begin{tabular}{@{}lc@{}}
\toprule
Fluence (neq/cm$^2$) & $\Delta R_{\mathrm{q}}/R_{\mathrm{q}}$ \\
\midrule
$3 \times 10^{11}$ & 2.5\% \\
$3 \times 10^{12}$ & 3.9\% \\
$6 \times 10^{12}$ & 0.8\% \\
$3 \times 10^{13}$ & 0.9\% \\
$6 \times 10^{13}$ & 9.4\% \\
\bottomrule
\end{tabular}
\end{table}

Two main features emerge from this analysis. First, the transition from $3 \times 10^{13}$ to $6 \times 10^{13}$\,neq/cm$^2$ results in an increase in $R_{\mathrm{q}}$ for 100\% of the devices, confirming the trend suggested by the median plots. Second, in approximately 67--78\% of the cases, $\Delta R_{\mathrm{q}}$ decreases between $3 \times 10^{12}$ and $6 \times 10^{12}$\,neq/cm$^2$.

Although a larger and more homogeneous sample would be required to accurately establish the functional dependence of radiation damage on fluence, comparing the median $\Delta R_{\mathrm{q}}$ with the corresponding quenching resistance indicates that irradiation below $6 \times 10^{13}$\,neq/cm$^2$ produces changes of only 1--4\%, comparable to the intrinsic device-to-device variability. At $6 \times 10^{13}$\,neq/cm$^2$, however, the average increase reaches approximately 9.4\%, suggesting the existence of a threshold between the two highest irradiation levels beyond which radiation damage becomes the dominant effect.

\begin{table}[htbp]
\centering
\caption{Percentage of sequences increasing (``Up'') in $\Delta R_{\mathrm{q}}$ between consecutive fluences.}
\label{tab:up-percent}
\begin{tabular}{@{}lccccc@{}}
\toprule
Transition & 297\,K & 223\,K & 173\,K & 123\,K & 78\,K \\
\midrule
$3 \times 10^{11} \to 3 \times 10^{12}$ & 77.8\% & 55.6\% & 55.6\% & 55.6\% & 55.6\% \\
$3 \times 10^{12} \to 6 \times 10^{12}$ & 22.2\% & 33.3\% & 33.3\% & 22.2\% & 33.3\% \\
$6 \times 10^{12} \to 3 \times 10^{13}$ & 44.4\% & 55.6\% & 44.4\% & 55.6\% & 66.7\% \\
$3 \times 10^{13} \to 6 \times 10^{13}$ & 100\%  & 100\%  & 100\%  & 100\%  & 100\%  \\
\bottomrule
\end{tabular}
\end{table}

\textbf{Outlying devices:} The response of the quenching resistance differs substantially among devices exposed to the same irradiation level. A qualitative examination identified four potential outliers:
\begin{itemize}
\item Device \#102 exhibits an average $\Delta R_{\mathrm{q}}$ of 111\,$\Omega$, approximately an order of magnitude larger than that of the other devices irradiated to the same fluence.
\item Device \#105 similarly shows consistently larger resistance shifts than the remainder of its group (143\,$\Omega$ versus 22 and 3\,$\Omega$). However, because the three devices irradiated at $3 \times 10^{12}$\,neq/cm$^2$ all display markedly different responses (medians of $\{23,\,132,\,0.4\,\Omega\}$), identifying the representative value for this group is not straightforward.
\item Device \#109 behaves differently from all other samples, exhibiting a systematic decrease in quenching resistance, with increasingly negative $\Delta R_{\mathrm{q}}$ values as the temperature decreases, reaching $-39$\,$\Omega$ at 78\,K. Nevertheless, the overall variation remains below 5\% of the absolute resistance, suggesting that this behavior may still be due to intrinsic device variability rather than radiation damage.
\item Device \#115 exhibits exceptionally large resistance shifts at $6 \times 10^{13}$\,neq/cm$^2$. Whereas the other two devices in the same irradiation group show average increases of approximately 45\,$\Omega$, device \#115 exhibits an average increase nearly six times larger.
\end{itemize}

Overall, these observations indicate that device-to-device variability is the dominant characteristic of the present sample and that, at least at low and intermediate irradiation levels, individual SiPMs may respond differently to radiation exposure.

\textbf{Dependence on Temperature at Constant Fluence:} The post-irradiation quenching resistance is considerably more sensitive to temperature than to irradiation level. Between 297\,K and 78\,K, $R_{\mathrm{q}}$ approximately doubles, whereas increasing the fluence from $3 \times 10^{11}$ to $6 \times 10^{13}$\,neq/cm$^2$ results in only a 10--20\% increase.

\begin{figure}[t]
  \centering
  \includegraphics[width=\linewidth]{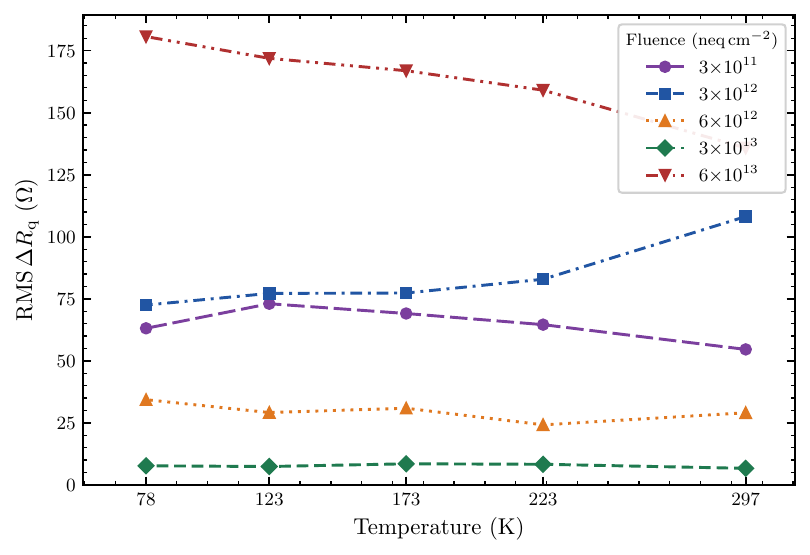}
  \caption{RMS of $\Delta R_{\mathrm{q}}$ as a function of operating temperature for each irradiation group. Unlike $\Delta V_{\mathrm{bd}}$, the radiation-induced resistance shift shows no systematic decrease with cooling.}
  \label{fig:rms_drq_temp}
\end{figure}

However, unlike the behavior observed for breakdown voltage, $\Delta R_{\mathrm{q}}$ does not decrease monotonically with decreasing temperature. Instead, as seen in Figure~\ref{fig:rms_drq_temp}, it fluctuates around nearly constant values, with only minor variations. A possible exception is the $6 \times 10^{13}$\,neq/cm$^2$ group, where $\Delta R_{\mathrm{q}}$ appears to increase slightly at lower temperatures, although this trend is not consistently observed for all devices.

Overall, $\Delta R_{\mathrm{q}}$ appears largely independent of temperature and remains comparable to --- or smaller than --- the standard deviation within each irradiation group, suggesting that much of the observed variation arises from intrinsic device-to-device differences. Moreover, since $R_{\mathrm{q}}$ nearly doubles while $\Delta R_{\mathrm{q}}$ remains approximately constant, the relative impact of radiation damage decreases with cooling.

\begin{table}[htbp]
\centering
\caption{Mean variation $\Delta R_{\mathrm{q}}$ after irradiation, and thermal coefficient $\mathrm{d}R_{\mathrm{q}}/\mathrm{d}T$.}
\label{tab:drq}
\begin{tabular}{@{}lcccccc@{}}
\toprule
Fluence (neq/cm$^2$) & 297\,K & 223\,K & 173\,K & 123\,K & 78\,K & $\mathrm{d}R_{\mathrm{q}}/\mathrm{d}T$ ($\Omega$/K) \\
\midrule
$3 \times 10^{11}$ & 54.6  & 64.6  & 69.1  & 73.0  & 63.1  & $-1.90$ \\
$3 \times 10^{12}$ & 108.1 & 82.9  & 77.3  & 77.2  & 72.5  & $-1.85$ \\
$6 \times 10^{12}$ & 29.1  & 24.2  & 30.9  & 29.2  & 34.4  & $-1.93$ \\
$3 \times 10^{13}$ & 6.7   & 8.3   & 8.5   & 7.4   & 7.7   & $-2.00$ \\
$6 \times 10^{13}$ & 135.8 & 159.1 & 166.9 & 171.9 & 180.7 & $-2.03$ \\
\bottomrule
\end{tabular}
\end{table}

Finally, the thermal coefficient $\mathrm{d}R_{\mathrm{q}}/\mathrm{d}T$ remains essentially unaffected by irradiation, showing only a slight increase above $6 \times 10^{12}$\,neq/cm$^2$ (see Table~\ref{tab:drq}). However, its magnitude increases nonlinearly during cooling, becoming approximately three times larger between room temperature and liquid-nitrogen temperature (from approximately $-1.1\,\Omega/\mathrm{K}$ to $-3.2\,\Omega/\mathrm{K}$), indicating an increasingly strong temperature dependence at lower temperatures.

\section{Discussion}

\subsection{Physical mechanisms of radiation damage}

Radiation damage in SiPMs arises primarily through two mechanisms. The first is bulk damage, associated with Non-Ionizing Energy Loss (NIEL), which results from neutron collisions with the silicon crystal lattice. These collisions displace silicon atoms from their lattice sites, generating vacancies and single-point and cluster defects within the crystal structure. The second mechanism is surface ionization damage, which occurs when radiation with energies below the threshold required to produce bulk damage (approximately 300\,keV) induces charge accumulation within the surface oxide (SiO$_2$) and at the Si--SiO$_2$ interface, modifying the internal electric field distribution of the device.~\cite{ref15}

The observed $V_{\mathrm{bd}}$ behavior is consistent with the expectations of bulk damage. High neutron fluences can reduce the effective doping concentration within the multiplication region, thereby increasing the voltage required to initiate avalanche breakdown~\cite{ref15,ref11}. In addition, the invariance of the thermal coefficient $\mathrm{d}V_{\mathrm{bd}}/\mathrm{d}T$ indicates that irradiation shifts the absolute value of $V_{\mathrm{bd}}$ without altering its intrinsic temperature dependence~\cite{ref16}.

$R_{\mathrm{q}}$, implemented using a complex structure of doped polycrystalline silicon, is instead more sensitive to local microstructural variations. Irradiation tends to increase the density of defects at grain boundaries, reducing effective carrier mobility and consequently increasing the electrical resistivity of the material~\cite{ref17}. This mechanism also amplifies the pre-existing device-to-device scatter, explaining the large dispersion observed in the post-irradiation response of $R_{\mathrm{q}}$. The thermal coefficient $\mathrm{d}R_{\mathrm{q}}/\mathrm{d}T$, like $\mathrm{d}V_{\mathrm{bd}}/\mathrm{d}T$, stays essentially unchanged.

\subsection{Implications for LHCb Upgrade II}

From the perspective of the LHCb Upgrade II, the most relevant operating conditions correspond to higher neutron-equivalent fluences of $3 \times 10^{13}$ and $6 \times 10^{13}$\,neq/cm$^2$ at cryogenic temperatures, which closely match the expected detector irradiation during the whole data taking, including a safety factor of 2.~\cite{ref7}

Under these conditions, a 0.91\,V shift in $V_{\mathrm{bd}}$ is non-negligible. Without a corresponding bias adjustment, the SiPM would operate at reduced over-voltage, approaching the sub-Geiger regime and losing substantial gain.

The observed reduction of $\Delta V_{\mathrm{bd}}$ at low temperatures is therefore an encouraging result. The post-irradiation bias correction shrinks, simplifying detector operation and calibration. Furthermore, the uniform response of the irradiated devices supports the feasibility of large-scale deployment from the standpoint of $V_{\mathrm{bd}}$ stability.

The implications of changes in quenching resistance are somewhat different. An increase in $R_{\mathrm{q}}$ leads directly to an increase in the quenching time, $\tau = R_{\mathrm{q}} \times C_{\mathrm{d}}$ and therefore the microcell dead time and the maximum achievable photon counting rate~\cite{ref13,ref15}. The present results suggest the existence of a radiation-damage threshold between $3 \times 10^{13}$ and $6 \times 10^{13}$\,neq/cm$^2$. Below it the $R_{\mathrm{q}}$ change stays comparable to intrinsic variability while above it rises substantially.

Unlike the behavior observed for the breakdown voltage, cooling the SiPMs does not appear to reduce the absolute radiation-induced variation in quenching resistance. At the highest fluence ($6 \times 10^{13}$\,neq/cm$^2$), the absolute shift may even increase slightly. Nevertheless, because the quenching resistance itself increases considerably as the temperature decreases, the relative impact of radiation damage is reduced under cryogenic operating conditions, representing a potential practical advantage.

On the other hand, the significant device-to-device variability observed in $R_{\mathrm{q}}$ remains a concern for detector operation. The lack of a uniform response among irradiated devices suggests that individual qualification and calibration may be required before installation in the final detector.

\subsection{Statistical limitations and future work}

The principal limitation of the present study is the small sample size ($n = 3$ devices per fluence), insufficient to separate genuine physical trends in $\Delta R_{\mathrm{q}}$ from statistical fluctuations. With only three devices, a single outlier can dominate the group statistics.

Repeating this study with a larger number of devices (6--10 SiPMs per irradiation level) would significantly improve the reliability of the observed trends of $R_{\mathrm{q}}$.

A larger dataset would enable more robust statistical analyses, including more rigorous identification of outliers, as well as more accurate estimates of group means, RMS deviations, and confidence intervals. Such an expanded study would also provide a stronger basis for determining whether the apparent radiation-damage threshold identified in this work represents a genuine physical phenomenon or is partly influenced by the limited statistical sample.

\section{Conclusions}

Fifteen Hamamatsu S13360 Silicon Photomultipliers (SiPMs) were characterised before and after neutron irradiation at five fluences ($3 \times 10^{11}$ to $6 \times 10^{13}$\,neq/cm$^2$) and five operating temperatures (room temperature to 78\,K). The objective of this study was to investigate the effects of neutron irradiation and operating temperature on the performance of these SiPMs, with regard to their $V_{\mathrm{bd}}$ and $R_{\mathrm{q}}$.

$V_{\mathrm{bd}}$ was shown to rise with fluence, with the magnitude of the radiation-induced shift becoming larger as the irradiation level increases. Below $3 \times 10^{13}$\,neq/cm$^2$ the shift stays within the intrinsic device spread, while at $6 \times 10^{13}$\,neq/cm$^2$ it reaches 0.91--1.15\,V and becomes statistically significant. Cryogenic operation reduces the damage by 20--40\,\% at the highest irradiation level. The thermal coefficient $\mathrm{d}V_{\mathrm{bd}}/\mathrm{d}T$ appears unaffected by irradiation and remains constant at $\sim$45\,mV/K.

$R_{\mathrm{q}}$ behaves more complexly because of greater sensitivity to the intrinsic device-to-device variability. It generally rises after irradiation; the increase becomes consistent for all temperatures at $6 \times 10^{13}$\,neq/cm$^2$, reaching $\sim$9.4\,\%. At lower fluences, the observed changes remain comparable to the intrinsic dispersion within the sample. The data therefore point to a radiation-damage threshold between $3 \times 10^{13}$ and $6 \times 10^{13}$\,neq/cm$^2$, above which radiation effects become significant and below which the intrinsic variability of the devices remains the dominant contribution.

The study also shows that $R_{\mathrm{q}}$ is considerably more sensitive to operating temperature than to irradiation. The quenching resistance approximately doubles when the temperature is reduced from 297\,K to 78\,K, whereas it increases by only about 20\% across the entire investigated fluence range. Although the radiation-induced variation $\Delta R_{\mathrm{q}}$ remains approximately constant with temperature, the substantial increase in the absolute value of $R_{\mathrm{q}}$ at cryogenic temperatures reduces the relative impact of radiation damage. Similarly, the thermal coefficient $\mathrm{d}R_{\mathrm{q}}/\mathrm{d}T$ appears largely independent of irradiation but increases significantly as the temperature decreases, approximately tripling between room temperature and liquid-nitrogen temperature (from about $-1.1\,\Omega/\mathrm{K}$ to $-3.2\,\Omega/\mathrm{K}$).

Overall, these results demonstrate that post-irradiation $V_{\mathrm{bd}}$ corrections can be reliably predicted from fluence and temperature because of the devices' uniform behavior. $R_{\mathrm{q}}$, by contrast, retains large device-to-device variability, and individual qualification and calibration may be required for high-radiation operation such as LHCb Upgrade II.

Furthermore, thanks to this study, a larger sample of custom SiPMs has already been ordered, with more stable values of $R_{\mathrm{q}}$ (metal quenching resistors instead of poly-silicon). Their characterization is planned for the coming months.

\addcontentsline{toc}{section}{Acknowledgements}
\section*{Acknowledgements}

I would like to express sincere gratitude to Dr.\ Wander Baldini for his invaluable guidance, continuous support, and availability throughout this study. I also gratefully acknowledge the Istituto Nazionale di Fisica Nucleare (INFN) and the University of Ferrara for providing the facilities and the opportunity to carry out this research.

\end{document}